# Hydrothermal liquefaction of sewage sludge; energy considerations and fate of micropollutants during pilot scale processing

Lars Thomsen[1,2], Pedro N. Carvalho[1,3], Juliano Souza Dos Passos[1,2], Konstantinos Anastasakis[1,2], Kai Bester[1,3], Patrick Biller[1,2*]

[1]WATEC – Centre for Water Technology, Aarhus University, Ny Munkegade 120, Aarhus 8000, Denmark

[2]Biological and Chemical Engineering, Aarhus University, Hangøvej 2, Aarhus 8200, Denmark

[3]Department of Environmental Sciences, Aarhus University, Frederiksborgvej 399, 4000 Roskilde, Denmark

**Abstract**

The beneficial use of sewage sludge for valorization of carbon and nutrients is of increasing interest while micropollutants in sludge are of concern to the environment and human health. This study investigates the hydrothermal liquefaction (HTL) of sewage sludge in a continuous flow pilot scale reactor at conditions expected to reflect future industrial installations. The processing is evaluated in terms of energy efficiency, bio-crude yields and quality. The raw sludge and post-HTL process water and solid residues were analyzed extensively for micropollutants via HPLC-MS/MS for target pharmaceuticals including antibiotics, blood pressure medicine, antidepressants, analgesics, x-ray contrast media, angiotensin II receptor blockers, immunosuppressant drugs and biocides including triazines, triazoles, carbamates, a carboxamide, an organophosphate and a cationic surfactant. The results show that a positive energy return on investment was achieved for all three HTL processing temperatures of 300, 325 and 350 °C with the most beneficial temperature identified as 325 °C. The analysis of the HTL by-products, process water and solids, indicates that HTL is indeed a suitable technology for the destruction of micropollutants. However, due to the large matrix effect of the HTL process water it can only be stated with certainty that 9 out of 30 pharmaceuticals and 5 out of 7 biocides products were destroyed successfully (over 98% removal). One compound, the antidepressant citalopram, was shown to be moderately recalcitrant at 300 °C with 87% removal and was only destroyed at temperatures ≥325 °C (>99% removal). Overall, the results suggest that HTL is a suitable technology for energy efficient and value added sewage sludge treatment enabling destruction of micropollutants.

1. **Introduction**

Hydrothermal liquefaction (HTL) is a promising technology for the production of advanced liquid biofuels and value-added chemicals through the conversion of a variety of organic materials. The concept involves the heating of aqueous slurries at high temperatures of around 300-350 °C, at pressures over the vapor pressure of water to avoid water vaporization and the accompanying latent heat of vaporization. The technology has been demonstrated for a variety of wet and dry feedstocks including lignocellulosic, algae and waste (Elliott et al. 2015a).

Lately, valorization of sewage sludge by HTL has been increasingly recognized and investigated (Chen et al. 2020). Sewage sludge can be found in urban areas as a waste stream from wastewater treatment plants (WWTPs). It represents a sustainable feedstock for HTL with low, or even negative value that shows a significant biomass potential. It's production in the EU27 is approximately 11.5 MT dry matter (DM) and is expected to increase to over 13 MT by 2020 (Verlicchi and Zambello 2015), in the US approximately 14 MT of dry sludge is produced per year (Skaggs et al. 2018). An assessment by Seiple et al. (2017) estimated that only around 50% of the sludge produced in the US is used beneficially which includes energy recovery through anaerobic digestion (AD) and land application of both post-AD and non-AD sludge as fertilizer (Seiple et al. 2017). The rest of the sludge is incinerated for safety reasons.

Primary and secondary sludge are usually combined for subsequent further treatment/utilization or disposal. Traditional disposal methods include landfill and incineration as well as land application for agriculture. However, environmental and circular economy issues have discouraged these applications. Most advanced WWTPs utilize anaerobic digestion (AD) for the further valorization and treatment of sewage sludge. AD is employed in order to reduce the organic content and volume of sewage sludge while at the same time recovering energy in the form of biogas (typically 60% $CH_4$, 40% $CO_2$). However, anaerobic digestion suffers from very long residence times of several days to weeks, requiring large digestion tanks and produces large volumes of digestate which still contains a significant proportion of carbon. Besides that, the use of digestate as fertilizer has been subject to controversy due to the presence of organic micropollutants (e.g. pharmaceuticals, personal care products, pesticides, hormones, etc.), which may have recalcitrant properties, becoming a hazard for the entire food chain. AD has been shown to achieve only moderate reductions in pharmaceuticals and estrogens (Malmborg and Magnér 2015).

Currently, the widespread presence of micropollutants (MPs) in aquatic systems is a major concern across the globe. About 143,000 organic compounds were registered in the European market in 2012; many of which would end up in water systems at some point in their lifecycle (Das et al. 2017). MPs are being detected in the environment at low ng/L to µg/L levels - mostly because WWTPs are not designed to remove these compounds (Ternes et al. 2004). Volatilization of MPs during any of the WWT steps is negligible due to their very low Henry's constant ($<10^{-5}$ atm-m$^3$/mol) [7]. The biodegradable compounds (e.g. ibuprofen or caffeine), are relatively easy to remove. However, a large fraction of MPs are not eliminated or biotransformed in traditional WWTPs, instead they end up adsorbed on suspended particulates, i.e., on primary, secondary sludge and digestate (Ben et al. 2018, Das et al. 2017).

HTL can offer a potential solution for tackling both the low carbon recovery and the moderate MPs elimination during AD of sewage sludge. This concept has been suggested as early as the 1980's when researchers from Japan published on the HTL of sewage sludge (Akira et al. 1986, Yokoyama et al. 1987), showing the conversion of mixed primary and secondary sludge to bio-crude. These early reports found that yields of up to 50% could be achieved on an organic basis. More recent publications have moved away from the slow heating rate batch reactors employed in the 80's to using high heating rate bomb-type reactors, more alike to modern continuous HTL reactors (Qian et al. 2017, 2020). For example (Madsen and Glasius 2019) reached a maximum bio-crude yield of 42.2 wt.% in batch reactors at 330 °C, 10 min, and 9 wt.% solid loading of sewage sludge. Marrone et al. (2018) recovered respectively 59, 39 and 49 % carbon from primary, secondary and tertiary sludge as bio-crude during continuous HTL (Marrone et al. 2018), while Biller et al. (2018) recovered 64% in batch HTL reactors from primary sludge (Biller et al. 2018). Subsequent upgrading of the sludge HTL bio-crude via hydrotreating produces renewable fuels with high proportion of middle distillate fuels diesel and kerosene (n-paraffins) and a high heating value (HHV) of around 45 MJ/kg (Castello et al. 2019) .

In addition, the harsh temperatures and pressures employed by HTL process can potentially aid to the removal of MPs present in sewage sludge. Currently the fate of such pollutants during HTL is largely unknown. To our knowledge there is only one study that has assessed the effect of HTL on bioactive compounds during HTL; Pham et al. (2013) investigated the fate of estrone (estrogen), florfenicol (antibiotic) and ceftiofur (antibiotic) during HTL in batch reactors at 250 and 300 °C and residences times of 15, 30 and 60 min (Pham et al. 2013). The results showed complete deactivation of antibiotic resistant genes and removal of bioactive compounds at reaction times over 15 min. These promising results motivated this study to assess the effect of HTL on the micropollutants found in real wastewater sludge at industrially relevant processing conditions.

The objective of the current paper is to study the HTL of sewage sludge at realistic conditions expected in industrial HTL plants in terms of residence time, heating rate and temperatures in continuous flow. The continuous flow experimental conditions employed, differ significantly to previously investigated batch reactors, where slow heating rates and long residence times (up to 60 mins) were used. The effect of temperature on the bio-crude yield and quality is assessed and used with processing data from the HTL pilot plant to optimize and evaluate the processing of sewage sludge to achieve a positive energy return on investment. At the same time, the possibility of removing MPs (pharmaceuticals and biocides) in real sewage sludge from an urban wastewater treatment plant is evaluated. MPs selection is based on existing methodologies used to study a wide array of chemical and biological water treatment processes and not necessarily a specific selection of representative compounds for HTL. Thus, unfortunately the range does not cover the compounds studied by Pham et al. (2013), but the list is presently much more extensive and comprising different types of pharmaceuticals, biocides and including some known transformation products (see **Table S1** and **Table S2** for chemical properties). This is the first time such an industrially realistic assessment on the fate of sewage sludge at different operating conditions and their effect on MPs and process efficiency has been carried out.

2. **Methodology**

    **2.1 Hydrothermal Liquefaction**

Approximately 4 $m^3$ of primary wastewater sludge were collected at the central wastewater treatment plant of the 40,000 inhabitants town Viborg located in Bruunshåb, Denmark (operated by Energi Viborg A/S). To increase the dry matter content, the sludge was filtered in a rotary vacuum drum filter, with a drum surface area of 1.83 $m^2$. The vacuum applied varied between 0.5-1 bar and a rotational speed of 0.5 rpm was employed. Previous results have shown negative energy return on investment (EROI) meaning more energy input into the system than the energy obtained in the main product when low DM slurries are subjected to HTL (Anastasakis et al.

2018). Higher DM slurries (approximately 20 wt.%) have been shown to increase bio-crude yields and improve the overall energy efficiency during HTL (Anastasakis et al. 2018).

During vacuum drum filtration samples of raw sludge, filtrate and filter cake were collected and analyzed. The sludge cake produced was transferred to a 1 m$^3$ paddle wheel mixer for homogenization and had its DM content reduced from 19 % to 16 % by addition of filtrate water to optimize its pumpability. Subsequently, the sludge cake was fed in the continuous HTL pilot reactor. The reactor has been previously described in detail (Anastasakis et al. 2018). Briefly; the reactor has a capacity of 19 L and consists of a 140 m length tubular system with a constant cross section with an internal diameter of 14.2 mm. The reactor employs a double pipe counter current heat exchanger where HTL products are used to pre-heat the influent slurry. A series of electric heaters are used to bring and maintain influent slurry to the desired temperature. Depressurization is achieved by a hydraulic take-off system consisting of two pistons of approximately 0.5 L capacity each, working in alternation. Products are collected and separated gravimetrically in a 90 L separation funnel after the separation of gaseous products by a cyclone. In this study, the reactor was operated for a total of 15 h, processing approximately 970 kg of slurry with a flow rate variating between, 39-94 L/h. Three different temperatures were investigated: 300, 325, and 350 °C. The temperature and residence time profiles of the three different temperatures investigated are depicted in **Figure 1** with the corresponding reactor sections indicated. Due to technical difficulties, the flow rate was larger during the 300 °C experimental window leading to a lower residence time compared to the other two experimental windows at 325 and 350 °C. The residence time, bio-crude yields, energy efficiency and EROI were determined using the methods described in detail previously (Anastasakis et al. 2018). Briefly, the residence time was calculated assuming that the change in the density of the slurry, due to different temperatures in the reactor, follows the density of water at different temperatures and pressures. The density of the feed slurry was calculated by taking into account the dry matter content of the slurry and by assuming a solids density of 500 kg/m$^3$. The thermal efficiency ($\eta_{th}$) represents the energy content of the bio-crude product compared to the energy content of the feedstock in terms of higher heating values (HHV) and product yield. The total energy efficiency of the system ($\eta_{tot}$) is analogue to $\eta_{th}$ with the addition of the external energy input to the system (in addition to the energy content of the feedstock) in terms of heating and electricity. Finally, EROI expresses the ratio of the energy content of the bio-crude product to the external energy supply to the main energy consuming units of the HTL system (trim heater, reactor and feed pump).

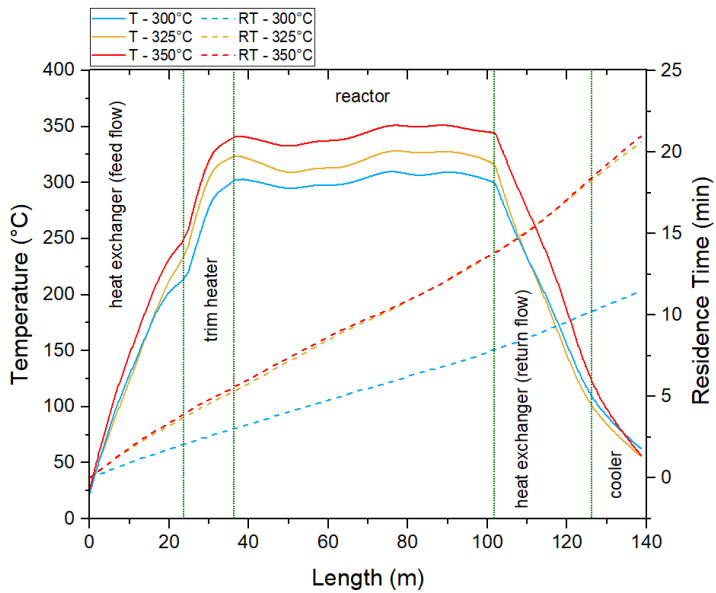

**Figure 1**: Temperature and residence time of the HTL reaction conditions investigated.

### 2.2 Product recovery and analysis

Once the reactor reached a steady state at the desired temperature, representative samples of bio-crude and process water were collected for further analysis, samples of approximately 2 kg were collected every 45 min. The process water was decanted from the bio-crude containing solids and portions of bio-crude from each temperature were dissolved in methanol and vacuum filtered through a Buchner filter setup using Whatman

Grade 1 cellulose filter papers. Methanol was removed using a rotary evaporator (Buchi R-300, Switzerland) to recover solids free bio-crude. The total amount of feedstock consumed, and bio-crude produced at steady state windows of several hours at 300, 325 and 350°C were considered to determine total bio-crude yields at a given temperature. The uncertainty in Table 2 is based on the deviation measured during the analyses of the samples (dry matter and solids content) and on the deviation measured by the HTL control system (flow rate and energy requirements).

Samples of sewage sludge, filtrate, filter cake, the HTL feed slurry and the HTL process water samples were dried in an oven at 105 °C for 24 h to measure the DM content. The dry samples were then heated to 550 °C for 5 h to determine the ash content. All the samples had their elemental composition analyzed using an Elementar vario MacroCube elemental analyzer (Langenselbold, Germany). The High Heating Value (HHV) of the samples were calculated using the equation proposed by Channiwala and Parikh (Channiwala and Parikh 2002) based on their elemental composition and ash content.

The moisture content of the bio-crude samples was determined by volumetric Karl fisher Titration (two component Hach-Lange KF1000). The ash content and elemental composition of the bio-crude samples were determined using the same procedure as mentioned before taking into account the water content. All measurements were repeated at least in duplicate.

HTL process water phase samples were analyzed for total organic carbon (TOC), COD, $NH_4$ and $PO_4$ content using Merck Spectroquant cuvette tests (part numbers: TOC-114879, COD-114541, $NH_4$-114559 and $PO_4$-114546)

Gas chromatography-mass spectrometry (GC-MS) was used to identify and quantify the volatile and semi-volatile components of the process HTL water. The procedure includes the derivatization of aqueous samples with methyl chloroformate (MCF). The method is described by Madsen et al. 2016 (Madsen et al. 2016). 200 µL of the process water samples were mixed with 40 µL NaOH (5%) and 40 µL pyridine, then derivatized by adding 2 x 20 µL MCF. The components were then extracted by adding 400 µL chloroform containing an internal standard (4-bromotoluene, 20 mg/L). To enhance the extraction and phase separation 400 µL NaHCO$_3$ (50 mM) were added. A total of 30 reference standards were prepared and derivatized using the same procedure for quantitative analysis making five-point calibration curves. The nonpolar phase was separated and analysed using an Agilent 7890B gas chromatograph coupled to an Agilent 5977A quadrupole mass spectrometer. The GC inlet was held at 280 °C with a split ratio of 20:1 and 1 µL was injected. A VF-5 ms column (60 m × 0.25 mm × 0.25 µm,) was used. The column oven was programmed to start at 60 °C, hold for 2 min, ramp at 5 °C/min to 200 °C, ramp at 20 °C/min to 320 °C and a final hold time of 5 min. The mass spectrometer transfer line and ion source were held at 300 °C. Electron impact ionization was employed at 70 eV and data was acquired in scan mode (35–400 m/z).

High performance liquid chromatography- mass spectrometry (HPLC-MS/MS) analysis were performed for the water-based samples and the process HTL solids to quantify the organic micropollutants present at the primary wastewater sludge and assess which substances could be found in the process HTL products. All samples were analyzed using two different HPLC-MS/MS methods, one to quantify biocides and the other for pharmaceuticals and in both cases selected transformation products. The target pharmaceuticals include antibiotics, blood pressure medicine, antidepressants, analgesics, x-ray contrast media, angiotensin II receptor blockers, immunosuppressant drugs and other pharmaceuticals commonly found in wastewater streams (chemical structures and properties are shown in **Table S1**). The target biocides include triazines, triazoles, carbamates, a carboxamide, an organophosphate and a cationic surfactant (chemical structures and properties are shown in **Table S2**), also found in wastewater streams. The set of target transformation products, selected based on their availability to the research group due to other ongoing research activities with ozonation and biological processes is presented together with the corresponding parent in the respective **Table S1** or **S2**.

All the samples for HPLC-MS/MS were kept at −20 °C until analysis. Before analysis, the samples were thawed at room temperature and vortex shaken for 30 s. Afterwards, 1.5 mL subsamples of each sample were transferred to an HPLC vial and centrifuged at 2500 RFC for 10 min. With a pipette, 1 mL of the supernatant was transferred to a new HPLC vial to which 100 µL of pharmaceuticals internal standard solution (**Table S3**) and 50 µL of biocides internal standard solution (**Table S4**) were added. Finally, 10 µL of the aqueous samples were injected into the HPLC system. The remaining supernatant in the centrifugation vial was discarded and the remaining pellet was further extracted. Solid samples, both the thawed solids, as well as the pellets were simply extracted by a two-step process, first water, second methanol, using an ultrasonic bath. Solid samples were weighted into 4 mL vials, 3 mL water was added and the vial was ultrasonicated for 15 min, followed by centrifugation (2500 RFC for 10 min) and further collection of the supernatant. Afterwards, the procedure was repeated but this second time using 3mL methanol. Pellets were extracted directly in the HPLC vial using the same approach, but in this case only 1.5mL of the solvent were used. With a pipette, 1 mL of the different supernatant was transferred to a new HPLC vial to which 100 µL of pharmaceuticals internal standard solution (**Table S3**) and 50 µL of biocides internal standard solution (**Table S4**) were added. Finally, 10 µL of the extracts were injected into the Agilent 1200 Series binary LC gradient HPLC system (Agilent Technologies, Santa Clara, CA, USA) which was coupled to a 5500 QTRAP (ABSciex, Framingham, MA, USA) triple quadrupole mass spectrometer. For both methods, a Synergy Polar RP column was used, and the gradient elution consisted of water and methanol, both containing 0.2% formic acid (v/v). The remaining methodological conditions were specific for each method and further detailed in the supplementary information. Analytical figures of merit are shown for the pharmaceuticals and biocides method in **Table S5** and **Table S6**, respectively. The concentration of the OMPs was calculated by linear regression of standard calibration (1/x weighting) by SCIEX OS v1.5 software. Concentrations reported for the water samples reflect the total content, meaning a sum of what was measured in the water phase and extracted from the respective pellet.

Removal rate was calculated based on the mass balance of the HTL system:

$$Removal\ (\%) = \frac{m_{in} - (m_{W,out} + m_{S,out})}{m_{in}} \times 100$$

Where, $m_{in}$ is the mass of pollutant fed to the system; $m_{W,out}$ is the mass of pollutant in the residual water phase; $m_{S,out}$ is the mass of pollutant in the residual solid phase. The amount of pollutant in the gas and bio-crude phase was considered negligible. When the compounds were not detected in the residual water or solid, the respective LOD was used to estimate the mass but if the compounds was detected then LOQ was used instead.

3. **Results and Discussions**

**3.1 Filtration process**

**Table 1** shows the DM and ash content of the filtration fractions of the vacuum drum filter. The untreated sewage sludge had a DM content of 3.5 wt.%. The drum filter generated a slurry cake with 19.3 wt.% DM, which was adjusted to 16 wt.% using the filtrate water to obtain a slurry which had optimal pumping performance in the HTL reactor. The vacuum drum filtration process was able to remove a large quantity of ash from the sludge. The initial ash content of 13 wt.% was reduced to 7 wt.% in the filter cake while it increased to 50 wt.% in the filtrate. This is most likely due to the presence of soluble salts in the liquid filtrate fraction. Reducing the ash content in the cake is beneficial for further processing as it increases the amount of organic material fed to the reactor per volume, makes product separation easier to perform and also results in lower ash content in the final bio-crude.

**Table 1**: analysis of sludge samples

|  | Sewage Sludge | ± | Cake / HTL feed slurry | ± | Filtrate | ± |
|---|---|---|---|---|---|---|
| Dry matter (wt.%) | 3.5 | 0.6 | 19.3[c] | 1.5 | 0.2 | 0.03 |
| C[a] (%) | 43.1 | 0.04 | 45.3 | 0.56 | 36.1 | 0.35 |
| H[a] (%) | 7.3 | 0.04 | 7.8 | 0.08 | 6.8 | 0.16 |
| N[a] (%) | 2.5 | 0.01 | 3.0 | 0.36 | 3.5 | 0.21 |
| S[a] (%) | 0.3 | 0.02 | 0.3 | 0.03 | 0.6 | 0.02 |
| O[a, b] (%) | 33.7 | 0.06 | 36.5 | 0.15 | 2.9 | 0.32 |
| Ash* (%) | 13.0 | 0.80 | 7.0 | 0.60 | 50.0 | 0.60 |
| HHV (MJ/kg)* | 19.9 | 0.07 | 21.1 | 0.13 | 19.3 | 0.35 |

[a]Dry matter based. [b]Calculated by difference. [c]Feed slurry dry matter was reduced to 16% for processing

### 3.2 Hydrothermal Liquefaction

#### 3.2.1 Bio-crude analysis

During the experimental campaign three different temperatures were investigated 300, 325 and 350 °C. A total of over 65 kg of bio-crude was produced and representative samples were taken from each temperature for further analysis. **Figure 2** shows the elemental analysis of the three bio-crudes with the calculated HHV and ash content. Due to the removal of solids from the bio-crude via filtration, the ash content is low ranging from 1 wt.% at 300 °C and 325 °C to 2 wt.% at 350 °C. The ash content of bio-crude is important to consider as it is undesirable for further processing especially for catalytic hydrotreatment and has to be removed either in the HTL reactor inline or post reaction. Marrone et al. (2018) for example, achieved an ash content from continuous processing of primary sludge of 0.38 wt.% using inline filtration (Marrone et al. 2018). A clear trend in increasing HHV can be observed with increasing temperature, resulting in a maximum at 350 °C of 34.5 MJ/kg, this is primarily due to the reduction of oxygen from 22.1 to 12.7 wt.%. There is an increase in nitrogen content observed with increasing temperature from 300 to 325°C from 2.5 to 3.2 wt.%, which is similar to the content at 350 °C. The nitrogen content observed in the bio-crudes are comparatively lower to e.g. (Conti et al. 2020) and (Marrone et al. 2018) who report nitrogen levels of 4.3-5.1 and 5.0-5.6 wt.% respectively. However, sourced wastewater treatment sludge for both cases also had higher amounts of nitrogen present in the feedstock 3.6-7.9 and 7.4 wt.% respectively compared to the 3.0 wt.% in the current study (See Table 1).The bio-crudes are comparable in oxygen contents to typical HTL bio-crudes produced from other biomasses which are commonly in the region of 10-15 wt.% (Madsen and Glasius 2019). The current results of around 14 wt.% oxygen for the 325 °C and 350 °C are within the upper range of published data from primary sludge batch and continuous processing (5-15wt.% oxygen) (Marrone et al. 2018, Mujahid et al. 2020, Qian et al. 2020). The results show that there is a combined heteroatom content of N and O between 24.6 and 15.7 wt.% which requires significant upgrading in a bio-crude post processing stage such as hydrotreating. Overall the bio-crudes are characterized by a high solid content before filtration (See **Table 2**), but equally have already been shown from previous campaigns to be suitable for further upgrading after ash removal via catalytic hydrotreatment to produce fuels with very low heteroatom content (Castello et al. 2019).

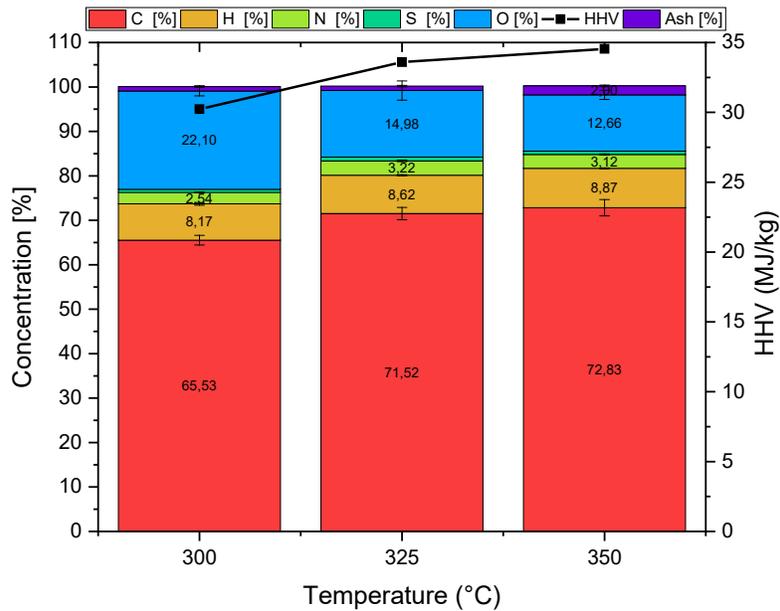

**Figure 2**: Elemental analysis and HHV of bio-crudes produced at different temperatures

### 3.2.2   HTL process energy evaluation

The continuous pilot scale HTL reactor allows constant monitoring of the energy input provided to the heating elements in the trim heater and the main reactor section. For each temperature a window of 1 h was defined to analyze the processing data and experimental results from bio-crude characterization. The results are summarized in **Table 2** for a representative 1 h window each. The energy in the feedstock is calculated from the flow rate, dry matter content of the slurry and the HHV of the dry sludge. The bio-crude yield ranged from 28 wt.% at 300 °C to a maximum of 64 wt.% at 325 °C. However, taking the solids and moisture content of the bio-crudes into account, the highest bio-crude yield of 41 wt.% was obtained at 325°C. This broadly confirms the results presented by Madsen and Glasius (2019) where the optimum processing conditions for primary sludge was identified in batch reactors as 330 °C, 10 min residence time yielding 42.2 wt.% bio-crude (Madsen and Glasius 2019). At 325 °C the bio-crude also exhibited a relatively high HHV of 33.6 MJ/kg which resulted in the highest thermal efficiency (chemical energy recovery) of 69%, in relation to the feedstock HHV and yield. Taking into consideration the amount of energy required to heat the reactor (7.5 kW) and pumping the slurry to the required pressure (0.4 kW) a total efficiency of 57% was achieved. When purely comparing the amount of energy produced as bio-crude (26.3 kW) to the input energy for heat and pumping, the energy returned on energy invested (EROI) for the reactor can be calculated. This results in a ratio of 3.4 for the 325 °C experiment meaning 3.4 times more energy is produced as bio-crude than is invested for processing. This ratio drops to 2.2 at 300 °C and 1.6 at 350 °C. In a previous publication using the same feedstock and reactor at 350°C but with a very low dry matter content of the slurry of only 4 wt.% vs. 16 wt.% in the current study, an EROI of 0.5 was achieved. This highlights the importance of high dry matter slurries for continuous HTL processing also due to the fact that the bio-crude yields are significantly higher with 53 wt.% vs 25 wt.% (prior solids removal) (Anastasakis et al. 2018). From a processing efficiency point of view, the data in **Table 2** suggests that processing at 325 °C is most suitable, even though the oxygen content in the bio-crude is slightly higher than at 350 °C. Due to the similar carbon content of the bio-crude at 325 and 350 °C but the greater yield at 325 °C there is more carbon found in the bio-crude fraction, which is consistent with the process water analysis presented in **Table 3**. It has to be highlighted that the flow rates in the reactor were not constant making the comparison of different temperatures difficult. However, flow rates and hence residence times were comparable for the experiments at 325 °C and 350 °C (43 and 39 L/h, respectively). Future campaigns will aim to highlight the effect of temperature and flow rate (residence time) independently by keeping one parameter constant. Nevertheless, the data shows that a positive EROI is possible at all three temperatures but considerable optimization opportunities exist by carefully adjusting the operating conditions to optimize the interchanging effects of temperature, residence time on the product yields and quality as well as the energy requirements for the reactor to operate. Finally, it should be noted that the EROI calculations only take into consideration the energy requirements of the HTL plant in terms of heat and pumping, not auxiliary equipment power, filtration of the sludge and any transportation of sludge.

Table 2: Energetic analysis of HTL sludge processing at different experimental conditions.

| Temperature | °C | 300 | ± | 325 | ± | 350 | ± |
|---|---|---|---|---|---|---|---|
| Flow rate | L/h | 94 | 0.3 | 43 | 0 | 39 | 0 |
| Dry matter content | (wt.%) | 16.0 | 1.4 | 16.0 | 1.4 | 16.0 | 1.4 |
| Feedstock consumed | (kg, dry) | 15.0 | 1.4 | 6.9 | 0.6 | 6.3 | 0.6 |
| Energy in Feedstock | (kW, dry) | 82.7 | 7.5 | 37.9 | 3.3 | 34.5 | 3.0 |
| Bio-crude yield (as received) | (wt.%) | 28.1 | 0.4 | 64.1 | 0.9 | 53.4 | 0.8 |
| Bio-crude yield (dry) | (wt.%) | 23.6 | 0.5 | 53.0 | 1.1 | 45.5 | 1.0 |
| Bio-crude yield (dry solids free) | (wt.%) | 17.7 | 0.5 | 40.8 | 1.2 | 31.8 | 0.9 |
| HHV bio-crude (dry solids free) | (MJ/kg) | 30.2 | 0.6 | 33.6 | 0.2 | 34.6 | 0.4 |
| Energy in Bio-crude (dry solids free) | (kW, dry) | 22.3 | 3.1 | 26.9 | 3.2 | 19.2 | 2.5 |
| Thermal efficiency ($\eta_{th}$) | (%) | 27.0 | 1.3 | 69.3 | 2.4 | 55.6 | 2.2 |
| Trim heater energy requirements | (kW) | 6.9 | 0.6 | 5.5 | 0.3 | 7.4 | 2.4 |
| Reactor energy requirement | (kW) | 2.3 | >0.01 | 2.0 | 0.3 | 4.2 | 0.1 |
| Main pump energy requirement | (kW) | 0.86 | NA | 0.40 | NA | 0.36 | NA |
| Total efficiency ($\eta_{tot}$) | (%) | 24.1 | 1.2 | 57.4 | 2.2 | 41.2 | 0.4 |
| EROI | | 2.2 | 0.2 | 3.4 | 0.2 | 1.6 | 0.1 |

### 3.2.3 HTL process water analysis

When implementing a HTL system for sludge valorization the fate of the resulting aqueous phase is an important factor to consider. As shown in **Table 3**, the COD concentrations of the samples range 40-50 g/L, which is two orders of magnitude higher than a typical wastewater treatment influent concentration (typically 200-800 mg/L). It is therefore troublesome to simply divert the HTL process water stream back to the WWTP influent. Although the volumes are relatively low compared to the overall inflow, the concentrations are very high, making other treatment options more attractive. The most commonly suggested options for this are anaerobic digestion and catalytic hydrothermal gasification (CHG). Both technologies have their advantages and disadvantages. AD suffers from typically long residence times, inhibition of the process by certain organics and low overall COD conversion (~60%) (Tommaso et al. 2015). Catalytic hydrothermal gasification has the advantage of low residence times, very high COD conversions (~99%) but suffers from the required catalyst being poisoned by sulphur (Elliott et al. 2015b, Marrone 2016). Both technologies also perform more efficiently at high carbon leadings due to the fact that more energy can be recovered from higher carbon concentrations per volume. The levels found in HTL process water of around 20 g/L are low if thermally efficient process water cleanup is to be performed by AD or CHG. The analysis of the process waters in **Table 3** shows similar compositions amongst the different temperatures. The TOC value is around 20% lower for the 325 °C produced water sample. This fits with the high bio-crude yields observed at this temperature where more carbon is fractionated to the bio-crude. The data further shows that the ammonium concentration increases with temperature as well as the N-containing pyrazine compounds. The employed GC-MS method is not complete and only the largest peaks were quantified when authentic standards were available. Other compounds that were present, exhibiting large peak areas, which are not quantified are 3-methoxy-pyridine and acetone. Overall, the data demonstrates the need to further cleanup of this process water while some of the identified compounds pose a threat to biological treatment due to the inhibitory or recalcitrant nature.

Table 3: Bulk analysis of HTL process waters.

| mg/L | 300 °C | ± | 325 °C | ± | 350 °C | ± |
|---|---|---|---|---|---|---|
| Total Organic Carbon (TOC) | 20,000 | 100 | 15,700 | 200 | 20,200 | 850 |
| Chemical Oxygen Demand (COD) | 50,833 | 438 | 42,250 | 250 | 51,850 | 50 |
| $NH_4^+$ | 373 | 3 | 635 | 5 | 720 | 5 |
| $PO_4^{3-}$ | 4.3 | NA | 2.3 | NA | 7.6 | NA |
| Acetic acid | 454.3 | 113.3 | 698.7 | 80.6 | 446.0 | 42.2 |
| Isobutyric acid | 14.4 | 1.3 | 13.9 | 1.9 | 15.0 | 3.8 |
| Pyrazine | 358.6 | 29.8 | 378.4 | 19.8 | 451.0 | 61.0 |
| Pyrazine, methyl- | 330.4 | 16.0 | 350.5 | 27.1 | 372.8 | 44.6 |
| 2-Cyclopenten-1-one, 2-methyl- | 55.6 | 8.7 | 73.6 | 6.9 | 63.1 | 6.2 |
| Pyrazine, 2,5-dimethyl- | 5.4 | 0.4 | 6.4 | 0.7 | 6.1 | 0.8 |
| Malonic acid | 13.6 | 1.6 | 10.4 | 6.5 | 1.4 | 0.1 |
| Pyrazine, ethyl- | 7.0 | 0.5 | 2.6 | 3.5 | 4.8 | 4.6 |
| Pyrazine, 2,3-dimethyl- | 5.8 | 0.6 | 7.1 | 0.9 | 6.8 | 1.1 |
| Methyl Malonic acid | 9.1 | 0.9 | 11.0 | 1.1 | 11.3 | 1.1 |
| 2-Cyclopenten-1-one, 3-methyl- | 140.6 | 16.5 | 213.8 | 20.5 | 219.9 | 25.4 |
| Levulinic acid | 19.2 | 2.0 | 16.2 | 2.7 | 18.7 | 2.7 |
| Pyrazine, trimethyl- | 4.3 | 0.3 | 5.4 | 0.5 | 4.9 | 0.5 |
| Succinic acid | 56.1 | 7.0 | 51.1 | 9.3 | 39.8 | 11.7 |
| Pyrrolidinone | 27.6 | 0.6 | 30.8 | 1.3 | 31.0 | 1.2 |
| 3-Hydroxypyridine monoacetate | 20.8 | 27.1 | 7.8 | 0.8 | 7.2 | 0.9 |
| Glutaric acid | | 7.9 | 0.3 | 8.9 | 0.9 | 7.2 | 0.8 |
| Phenol | | 7.3 | 1.0 | 7.0 | 0.9 | 6.0 | 1.2 |
| Pyrocatechol | | 71.8 | 16.2 | 57.6 | 0.4 | 57.4 | 0.6 |

### 3.3 Organic micropollutants

The results from both the pharmaceuticals (**Table S7**) and the biocides (**Table S8**) analysis indicate similar behavior for both type of organic micropollutants. The overall trend indicates the presence of several compounds in the sewage sludge fed to the HTL system as well as on the respective drum filtrate water and slurry fed to the HTL. Regarding the pharmaceuticals, 11 of the 36 parent compounds analyzed and 12 of the 16 known transformation products were not detected in the sewage sludge and HTL slurry. Thus, a total of 25 pharmaceuticals (azithromycin, clarithromycin, erythromycin, roxithromycin, ciprofloxacin, clindamycin, sulfamethizole, trimethoprim, atenolol, metoprolol, propranolol, carbamazepine, citalopram, diclofenac, ibuprofen, tramadol, venlafaxine, iomeprol, valsartan, irbesartan, candesartan, mycophenolic acid, oxazepam, benzotriazole and gabapentin) and 4 transformation products (alpha-hydroxymetoprolol, *O*-demethylmetoprolol, (10S,11S)-dihydroxy-10,11-dihydrocarbamazepine and diclofenac amide) were present with quantifiable levels in the slurry fed to the HTL. Two of the x-ray contrast media, iohexol and iopromide, known as hydrophilic could only be determined in the drum filtrate but not on the sludge/slurry. Concentrations in the sludge and HTL slurry ranged from LOQ (clindamycin, *O*-demethylmetoprolol) up to 131 µg/kg for losartan, in line with the levels (0.01 – 5000 µg/kg) reported in the literature for the different compounds (Lindholm-Lehto et al. 2018).

Out of 16 biocides tested, only 7 were detected and/or quantified, tebuconazole, carbendazim, diflufenican and its transformation product (TP) 2-[3-(trifluoromethyl)phenoxy]pyridine-3-carboxylic acid (known as AE-B) and the cationic surfactant benzalkonium chloride (also known as alkyldimethylbenzylammonium chloride / ADBAC) in its different even-numbered alkyl chain lengths 12-BAC, 14-BAC and 16-BAC. Concentrations were very different between the type of compounds. Tebuconazole and carbendazim were in similar ranges (0.1 to 10 µg/kg) to those reported previously for sewage sludge (Chen et al. 2012, Kupper et al. 2006). AE-B was only detected in the different samples, and diflufenican was in similar levels to tebuconazole and carbendazim. For the best of our knowledge, diflufenican has been reported previously in streams and potentially linked with effluent discharges (Muenze et al. 2017) but never directly quantified in sewage sludge. The levels of benzalkonium chlorides were much higher (100 – 5000 µg/kg) than for the other biocides. Benzalkonium chlorides belong to the broader family of quaternary ammonium compounds, a major class of cationic surfactants, used as the ingredients in fabric softeners, antistatics, disinfectants, biocides, detergents, phase transfer agent and numerous personal care products (Zhang et al. 2015). Benzalkonium chlorides are known to sorb to sludge in wastewater treatment plants and the values presently found are within the orders of magnitude previously reported (µg to mg per kg of sludge) (Zhang et al. 2015). The analytical method was not extensively optimized for the solids extraction thus the values shown should not be considered as absolute but indicative of their order of magnitude.

After the HTL treatment both residual water and solids were analyzed. The HTL process water, most probably due to the higher organic content than typical wastewater, presented a significant matrix effect leading to significant ion suppression for some of the pharmaceutical and biocide compounds. It should be noted that this ion suppression effect is not observed for typical wastewater (Table S5), and was not observed for the extracts from the HTL solids. Therefore, for the compounds for which isotope dilution measurements are possible due to availability of the respective internal standard, the removal is unquestionable (**Table 4**). In addition, the HTL solids revealed no detectable level of compounds for both pharmaceuticals and biocides. It can, thus, be clearly stated that the HTL was capable of removing 9 out of the 30 pharmaceutical compounds measured in the HTL feed: azithromycin 99.8%, clindamycin 83.7 – 84.8%, atenolol 98.8 – 98.9%, citalopram 86.7 – 99.9%, diclofenac 98.7-98.8%, ibuprofen 98.4 – 98.5%, iomeprol 99.7%, losartan 99.9% and mycophenolic acid 99.8%. Regarding the biocides, 5 out of the 7 compounds were also effectively removed: tebuconazole 99.3 – 99.4%, diflufenican 99.9%, 12-BAC 99.9%, 14-BAC 99.9% and 16-BAC 99.9%. Removals above 98% were the most common. The exception was clindamycin that was only detected in the feed (thus the removal rate calculation is only based on the limits of the analysis, not necessarily of the process) and citalopram that presented a removal rate dependent of the temperature. We presently demonstrate that several other micropollutants can be removed above 98%, besides the previously studied estrone, florfenicol and ceftiofur [13]. It should be noted that HTL presents high removal for compounds that are usually recalcitrant to other treatment processes: for instance, diclofenac presents very limited removal by biological processes and requires advanced oxidation processes (Beltrán et al. 2009) or ibuprofen that is easily biodegraded but recalcitrant to ozonation (Quero-Pastor et al. 2014).

Carbamazepine, tramadol, venlafaxine, valsartan, irbesartan, candesartan, oxazepam, gabapentin, alpha-hydroxymetoprolol, *O*-demethylmetoprolol, (10S,11S)-dihydroxy-10,11-dihydrocarbamazepine and diclofenac amide) seem to be also removed in the HLT, but lack confirmation by an internal standard. The compounds that were affected by the ion suppression problem (clarithromycin, erythromycin, roxithromycin, ciprofloxacin, sulfamethizole, trimethoprim, metoprolol, propranolol, and benzotriazole) in the process water matrix did not reveal any peak, but conclusions cannot be drawn. For the biocides, carbendazim suffered ion suppression in the HTL water samples, thus it can only be confirmed that it is not present in the HTL solids after the treatment. AE-B seems to be totally removed, the compound was not detected in either the HTL process water or solids, but lacks confirmation by an internal standard.

A very limited number of compounds seem to be partly recalcitrant to HTL process. Citalopram was quantified in the HTL water from the 300 °C process, but not for the higher temperatures (325 and 350 °C). The same seems true for benzotriazole and the transformation product alpha-hydroxymetoprolol. These findings suggest that 300

°C might not be sufficiently high temperature or/and the residence time was too short (approximately 10 min RT at 300 °C compared to 20 min at 325 °C and 350°C) for the efficient elimination of the aforementioned compounds. Further attention should be paid to the formation of transformation products. We observed the potential formation of 5-amino-3-methylisoxazone in the process water, not fully confirmed by an internal standard and not easy to explain once the respective parent compound sulfamethoxazole was not measured in the feed (Table S7). Nevertheless, isoxazole rings (azole, five-membered nitrogen heterocyclic ring, with an oxygen atom next to the nitrogen) are commonly found in a number of antibiotics and pesticides (Chen et al. 2018) thus, other potential parent compounds might be present in the feed. It should be stressed that the current targeted transformation products are based on the pre-existing knowledge from other type of chemical and biological systems. Thus, there is a need for future studies to clarify the removal pathways and potential formation of transformation products in HTL.

All in all, the HTL technology reveals a big potential to cope with the contamination by organic micropollutants.

**Table 4:** Removal rate of pharmaceuticals and pesticides in the hydrothermal reactor at the different temperatures, based on the mass balance of the system

| Group | Compound | HTL 300 °C Removal % | ± | HTL 325 °C Removal % | ± | HTL 350 °C Removal % | ± |
|---|---|---|---|---|---|---|---|
| Antibiotics | Azithromycin | 99.8 | <0.1 | 99.8 | <0.1 | 99.8 | <0.1 |
| | Clindamycin | 84.3 | <0.1 | 85.4 | 0.1 | 85.3 | <0.1 |
| Blood pressure regulators | Atenolol | 98.8 | <0.1 | 98.9 | <0.1 | 98.9 | <0.1 |
| Antidepressant / Analgesic | Citalopram | 86.7 | <0.1 | 99.9 | <0.1 | 99.9 | <0.1 |
| | Diclofenac | 98.7 | <0.1 | 98.8 | <0.1 | 98.8 | <0.1 |
| | Ibuprofen | 98.4 | 0.1 | 98.5 | 0.1 | 98.5 | 0.1 |
| X-ray contrast media | Iomeprol | 99.7 | <0.1 | 99.7 | <0.1 | 99.7 | <0.1 |
| Angiotensin II receptor blocker | Losartan | 99.9 | <0.1 | 99.9 | <0.1 | 99.9 | <0.1 |
| Other uses | Mycophenolic acid | 99.8 | <0.1 | 99.8 | <0.1 | 99.8 | <0.1 |
| Triazoles | Tebuconazole | 99.3 | 0.1 | 99.4 | 0.1 | 99.3 | 0.1 |
| Carbamates | Carbendazim | 98.8 | <0.1 | 98.9 | <0.1 | 98.9 | <0.1 |
| Carboxamide | Diflufenican | 99.9 | <0.1 | 99.9 | <0.1 | 99.9 | <0.1 |
| Cationic surfactant | 12-BAC | 99.9 | <0.1 | 99.9 | <0.1 | 99.9 | <0.1 |
| | 14-BAC | 99.9 | <0.1 | 99.9 | <0.1 | 99.9 | <0.1 |
| | 16-BAC | 99.9 | <0.1 | 99.9 | <0.1 | 99.9 | <0.1 |

## 4. Conclusions

Overall this work has shown that hydrothermal liquefaction is a suitable technology for beneficial use of sewage sludge due to several advantages that can be concluded from this study:

- A positive energy return on investment is possible by processing sewage sludge at sufficient dry matter content with modern continuous flow HTL reactors, producing around three times more energy in the bio-crude than is required for pumping and heating.
- The highest bio-crude yields are obtained at 325 °C, broadly agreeing with batch literature data. At this temperature the highest EROI is also achieved, while the bio-crude quality is slightly superior at 350 °C.
- The process waters are very high in COD (~50 g/L), requiring further treatment and also introducing a large matrix effect for organic micropollutants analysis.

- At least 9 out of 30 pharmaceuticals and 5 out of 7 biocides are removed with over 98%. – Whether other compounds are being removed needs to be tested after analytical method adaption/cleanup.

5. Acknowledgments

This research was funded by the European Union's Horizon 2020 research and innovation program under grant agreement No. 764734 (HyFlexFuel—Hydrothermal liquefaction: Enhanced performance and feedstock flexibility for efficient biofuel production) and the European Research Council (ERC) under the European Union's Horizon 2020 research and innovation program grant No. 849841 (REBOOT- Resource efficient bio-chemical production and waste treatment).